\documentstyle[12pt]{article}

\newcommand{\sect}[1]{\setcounter{equation}{0}\section{#1}}

\def\be{\begin{equation}}
\def\ee{\end{equation}}
\def\bea{\begin{eqnarray}}
\def\eea{\end{eqnarray}}

\newfont{\frak}{eufm10 scaled\magstep1}
\newcommand{\fra}[1]{\mbox{\frak #1}}

\font\black=msbm10 scaled\magstep1
\def\extra #1{\hbox{{\black #1}}}
\def\R{{\hbox{{\extra R}}}}

\def\k{\kappa}
\def\w{\omega}
\def\om{\Omega}
\def\te{\theta}

\def\hhp{H^{(1)}}

\def\xxp{{\cal S}^{(1)}}
\def\xxl{{\cal S}^{(2)}}

\def\Sk{{\,\mbox{S}}}           
\def\Ck{{\,\mbox{C}}}

\def\diag{\,\mbox{diag}\,}

\def\>#1{{\bf #1}}                 
\def\1{\'{\i}}                           
\def\back{\!\!\!\!\!}

\begin{document}
 
\thispagestyle{empty}

\hfill\quad\ 
\ 
\vspace{2cm}

\begin{center}

{\LARGE{\bf{Homogeneous phase spaces:}}} 

 {\LARGE{\bf{the Cayley--Klein framework}}} 

\end{center}

\bigskip

\begin{center}
Francisco J. Herranz  

{\em { Departamento de F\1sica, E. U. Polit\'ecnica, }
\\  Universidad de Burgos, E-09006, Burgos, Spain \\ 
email: fteorica@cpd.uva.es}
 \end{center}

\begin{center}
Mariano Santander 

{ \em {   Departamento de F\1sica Te\'orica, Facultad de Ciencias, } \\ 
 Universidad de Valladolid,  E-47011, Valladolid, Spain \\ 
email: santander@cpd.uva.es}
\end{center}
 
\bigskip

\begin{abstract} 
The metric structure of homogeneous spaces of rank-one and rank-two
associated to the real pseudo-orthogonal groups $SO(p,q)$ and some of their
contractions (e.g.,  $ISO(p,q)$, Newton--Hooke type groups\dots) is studied.
All these spaces are described from a unified setting following a
Cayley--Klein scheme allowing to simultaneously study  the main features of
their  Riemannian, pesudoRiemannian and semiRiemannian metrics, as well as
of their curvatures. Some of the rank-one spaces are naturally interpreted
as spacetime models. Likewise, the same natural interpretation for rank-two 
spaces is as  spaces of lines in rank-one spaces; through this relation
these rank-two spaces  give rise to homogeneous phase space models. The main
features of the phase spaces for homogeneous spacetimes are analysed.
\end{abstract}



\sect{Introduction}

The main aspect usually considered when working on a phase space is  its
associated symplectic structure. However {\it homogeneous} phase spaces,
i.e., those admitting a structure preserving a Lie group of transformations
have a richer structure, which can be overlooked if attention is focused
only on their symplectic structure. Actually, homogeneous phase spaces have
also a canonical connection and a metric structure, with a `main' Riemannian
metric (which can be as well pseudoRiemannian or degenerate Riemannian). In
some cases, the phase spaces have also invariant foliations, with a
subsidiary metric defined in each leaf.

The aim of this paper is to provide a complete characterization of the
quadratic metric in a set of phase spaces which are constructed as
symmetrical homogeneous spaces coming from the orthogonal Cayley--Klein (CK)
groups
\cite{tesis,conform}. This family of real Lie groups, called
`quasi-orthogonal'
\cite{Ros} are exactly the family of motion groups of the geometries of a
real space  with a projective metric
\cite{Sommer,Yas}. They include the semisimple pseudo-orthogonal  groups of
the Cartan series $B_l$, $D_l$ as well as many others which are
non-semisimple and that  can be obtained by contraction processes from the
formers (e.g., Euclidean, Poincar\'e, Galilean, Newton--Hooke type
groups\dots).  All groups in this family  share many important properties
allowing their study to be done at once.  Furthermore,  the kinematical
groups associated to different homogeneous models of spacetime \cite{BLL}
belong to this family, and this fact indeed provides  one of the strongest
physical motivations to study them.  

We first introduce in Section 2 the family of orthogonal CK groups from an
`abstract' point of view. In Section 3 we focus on the set of rank-one
homogeneous spaces associated to this family, and we describe in detail how
the metric(s) in these spaces comes from the Killing--Cartan form.
Physically, all homogeneous models of spacetime are rank-one spaces. In
Section 4 we carry out a similar study for the rank-two spaces; this is also
physically meaningful, because homogeneous phase spaces are rank-two spaces.
The last Section is devoted to commenting upon the physical meaning of
properties of the homogeneous phase spaces, which are presented against the
bakground provided by the more familiar properties of spacetime models.


\sect{Symmetrical homogeneous CK spaces}

The  orthogonal CK algebras are real Lie algebras of dimension $N(N+1)/2$ 
whose generators are  $\om_{ab}$ with $a,b=0,1,\dots, N$ and   $a<b$. This
family can be described collectively by means of $N$  real coefficients
$\w_1,\dots,\w_N$. The non-zero Lie brackets are  (no sum over repeated
indices):
\be
[\om_{ab}, \om_{ac}] =  \w_{ab}\om_{bc}   \qquad
[\om_{ab}, \om_{bc}] = -\om_{ac}   \qquad
[\om_{ac}, \om_{bc}] =  \w_{bc}\om_{ab}    
\label{ca}  
\ee
with $a<b<c $, and where the coefficients with two indices are defined by
\be
\w_{ab}:=\w_{a+1}\w_{a+2}\cdots\w_b\qquad a,b=0,1,\dots,N 
\qquad  a<b   \label{cb}
\ee
 satisfying
\be 
\w_{ac}=\w_{ab}\w_{bc}  \qquad \w_{a}=\w_{a-1\, a} .
\label{cc} 
\ee

We will denote ${\fra {so}}_{\w_1,\dots,\w_N}(N+1)$ the generic algebra in
the orthogonal CK family \cite{tesis,conform}. These algebras have a
fundamental  or {\it vector} representation by $(N+1)\times (N+1)$ real
matrices 
\be
\om_{ab} \to -\w_{ab}e_{ab}+e_{ba}
\label{cd}
\ee
where $e_{ab}$ is the matrix with a single   non-zero entry, 1, in the row
$a$ and column $b$. By exponentiation this representation allows to define
the orthogonal CK groups denoted $SO_{\w_1,\dots,\w_N}(N+1)$, whose 
one-parameter subgroups are easy to compute:
\be
e^{x \om_{ab}} =
   {\sum_{{\scriptstyle s=0\atop \scriptstyle s\ne a,b}}^N} e_{ss} + 
   \Ck_{\w_{ab}}(x) (e_{aa}+e_{bb}) +
   \Sk_{\w_{ab}}(x)( -\w_{ab} e_{ab}+e_{ba}) 
\label{ce}
\ee
where we introduce the `labeled' cosine $\Ck_\w(x)$ and sine
$\Sk_\w(x)$ functions defined by \cite{CKdosPoisson}:
\be 
\Ck_{\w}(x) = \left\{
\begin{array}{ll}
  \cos {\sqrt{\w}\, x} &\   \w >0 \cr 
  1  &\  \w  =0 \cr
\cosh {\sqrt{-\w}\, x} &\  \w <0 
\end{array}
\right. 
\quad
\Sk_{\w}(x) = \left\{
\begin{array}{ll}
    \frac{1}{\sqrt{\w}} \sin {\sqrt{\w}\, x} &\  \w >0 \cr 
  x &\  \w  =0 \cr 
\frac{1}{\sqrt{-\w}} \sinh {\sqrt{-\w}\, x} &\  \w <0 
\end{array}
\right. 
\label{bh}
\ee

Each coefficient $\w_a$ can be   scaled to the values $+1$, $0$, $-1$, thus
the family $SO_{\w_1,\dots,\w_N}(N+1)$ essentially  comprises $3^N$ Lie
groups; we remark that a value of any coefficient $\w_a$ equal to zero is
equivalent to a contraction limit. The properties of groups in the CK family
depend mainly on whether the  values of the coefficients $\w_a$ in the
sequence $(\w_1,\dots,\w_N)$ are equal to zero or not. The essential
properties are \cite{tesis}:   
 
\noindent
$\bullet$ When all $\w_a\ne 0$ $\forall a$, the group
$SO_{\w_1,\dots,\w_N}(N+1)$ is isomorphic to a semisimple pseudo-orthogonal
group  $SO(p,q)$ ($p+q=N+1$) in the Cartan series $B_l$ and $D_l$. Their
vector matrix representation (\ref{cd}) acting in $\R^{N+1}$ via matrix
multiplication leaves invariant a quadratic form whose matrix is  
$\Lambda_0^{(1)}=\diag(1,\w_{01},\w_{02},\dots,\w_{0N})$, so the values
$(p,q)$ can be determined as the number of positive and negative terms in
this sequence.
  
\noindent
$\bullet$ When a constant $\w_a = 0$, the group
$SO_{\w_1,\dots,\w_{a-1}, \w_a=0, \w_{a+1}, \dots, \w_N}(N+1)$ has a
semidirect structure:
$$
SO_{\w_1,\dots,\w_{a-1}, \w_a=0, \w_{a+1}, \dots ,\w_N}(N\! +\! 1)
\equiv T \odot   ( 
SO_{\w_1,\dots,\w_{a-1}}(a) \otimes
SO_{\w_{a+1}, \dots, \w_N}(N \!+ \!1 \!-\! a)) 
$$ 
where $T$ is an abelian subgroup of dimension $a(N+1-a)$. 

Repeated application of these results leads to an explicit  description of
the structure of CK groups according to the values of the constants $\w_i$:

\noindent
$\bullet$ Only $\w_1=0$, other different from zero.  We find the
inhomogeneous groups with a semidirect product  structure:
$$
SO_{0,\w_2,\dots,\w_N}(N+1)
\equiv T_N\odot  SO_{\w_2,\dots,\w_N}(N)\equiv ISO(p,q)  \qquad
p+q=N .
$$
The Abelian subgroup is $T_N$ generated by $\langle \om_{0b};\
b=1,\dots,N\rangle$ and 
$SO_{\w_2,\dots,\w_N}(N)$ is a pseudo-orthogonal group which preserves a
quadratic form whose matrix is $\diag (+,\w_{12},\dots,\w_{1N})$. The
Euclidean group $ISO(N)$ appears in this case when $(\w_1,\w_2,\dots,
\w_N)={(0,+,\dots,+)}$; the Poincar\'e group   $ISO(N-1,1)$ is reproduced
several times, e.g. for $(0,-,+,\dots,+)$,  $(0,+,\dots,+,-)$, etc.
 
\noindent
$\bullet$ $\w_1=\w_2=0$, other different from zero. Here we have  two
different semidirect structures for the CK group.  The one associated with
the vanishing of $\w_1$ is:
$$
SO_{0,0,\w_3,\dots,\w_N}(N+1) \equiv T_N \odot  
SO_{0,\w_3,\dots,\w_N}(N)
$$
where the second factor has again a semidirect  structure due to the
vanishing of
$\w_2$:
$$ 
SO_{0,0,\w_3,\dots,\w_N}(N+1) \equiv T_N \odot  \left( T_{N-1}\odot  
SO_{\w_3,\dots,\w_N}(N-1)\right) \equiv IISO(p,q)
$$
with  $p+q=N-1$. The alternative semidirect structure can be written
similarly. The Galilean group $IISO(N-1)$ appears in this  case  associated
to $(0,0,+,\dots,+)$.

\noindent
$\bullet$ $\w_a=0$, $a\notin\{ 1,N\}$.  These groups have a structure
$T_{a(N+1-a)} \odot (SO(p,q) \otimes SO(p',q'))$   \cite{WB}.  In
particular, for $\w_2=0$ we have
$T_{2N-2} \odot (SO(p,q)\otimes SO(p',q'))$ with $p+q=N-1$ and $p'+q'=2$,
which include for $q=0$ the oscillating  and  expanding  Newton--Hooke
groups \cite{BLL} associated to
$(+,0,+,\dots,+)$ and $(-,0,+,\dots,+)$, respectively.  
 
\noindent
$\bullet$ The extreme contracted case in the CK family corresponds to
setting all constants  $\w_a=0$. This is the so-called flag space group 
$SO_{0,\dots,0}(N+1)\equiv I\dots ISO(1)$ \cite{Ros}. In our notation it
should be understood $ISO(1)\equiv\R$.

The CK algebra ${\fra {so}}_{\w_1,\dots,\w_N}(N+1)$ can be endowed with an
Abelian group ${\extra Z}_2^{\otimes N}$ of involutive automorphisms
 generated by $N$ involutions: $\Theta^{(1)},\dots,
\Theta^{(N)}$. The action of $\Theta^{(m)}$ on the generators $\om_{ab}$ is
as follows:
\be
\Theta^{(m)}(\om_{ab})=
\left\{
 \begin{array}{rl}
\om_{ab}&\mbox{if either $a\ge m$ or $b<m$}\cr 
-\om_{ab}&\mbox{if   $a<m$ and $b\ge m$} 
\end{array}\right.
\label{da}
\ee
Each involution $\Theta^{(m)}$ provides a Cartan-like decomposition of
the CK algebra in antiinvariant  and invariant subspaces, denoted
${\fra p}^{(m)}$ and ${\fra h}^{(m)}$, respectively:
\be
{\fra {so}}_{\w_1,\dots,\w_N}(N+1)={\fra p}^{(m)}\oplus {\fra h}^{(m)}.
\label{db}
\ee
The set ${\fra h}^{(m)}$ of invariant elements is a Lie subalgebra, with a
direct sum structure:
\be
{\fra h}^{(m)}={\fra {so}}_{\w_1,\dots,\w_{m-1}}(m)\oplus
{\fra {so}}_{\w_{m+1},\dots,\w_N}(N+1-m),
\label{dc}
\ee
while the vector subspace ${\fra p}^{(m)}$ is not always a subalgebra. The
decomposition (\ref{db}) can be graphically visualized by arranging the
generators of ${\fra {so}}_{\w_1,\dots,\w_N}(N+1)$ in the form  of a 
triangle

\medskip

\noindent\hskip 1truecm
\begin{tabular}{cccc|cccc}
$\om_{01} $&$ \om_{02} $& $ \ldots $&$ \om_{0\,m-1} $&$ \om_{0m} $
&$ \om_{0\,m+1}$& $ \ldots $&$ \om_{0N} $  \\
 &$ \om_{12} $& $ \ldots $&$ \om_{1\,m-1} $&$ \om_{1m} $
&$ \om_{1\,m+1}$& $ \ldots $&$ \om_{1N} $  \\
 & & $ \ddots $&$ \vdots $&$ \vdots$
&$  \vdots$& $   $&$ \vdots $  \\
 & &  &$ \om_{m-2\,m-1} $&$ \om_{m-2\,m} $
&$ \om_{m-2\,m+1}$& $ \ldots $&$ \om_{m-2\,N} $  \\
 & &  & &$ \om_{m-1\,m} $
&$ \om_{m-1\,m+1}$& $ \ldots $&$ \om_{m-1\,N} $  \\
\cline{5-8}
 & &  &   \multicolumn{2}{c}{\,}
&$ \om_{m\,m+1}$& $ \ldots $&$ \om_{m\,N} $  \\
 & &  & \multicolumn{2}{c}{\,}
& & $ \ddots $&$ \vdots $  \\
 & &  & \multicolumn{2}{c}{\,}
& &  &$ \om_{N-1\,N} $
\end{tabular}

\medskip

\noindent
The subspace
${\fra p}^{(m)}$ is spanned by those $m(N+1-m)$ generators inside the
rectangle; the  left and down
triangles  correspond  respectively to the subalgebras 
${\fra {so}}_{\w_1,\dots,\w_{m-1}}(m)$ and 
${\fra {so}}_{\w_{m+1},\dots,\w_N}(N+1-m)$ of    ${\fra h}^{(m)}$. 

Relative to the decomposition (\ref{dc}), the structure of commutators
(\ref{ca}) in the CK algebra can be summed up as:
\be
[{\fra h}^{(m)}, {\fra h}^{(m)}] \subset {\fra h}^{(m)}  \qquad 
[{\fra h}^{(m)}, {\fra p}^{(m)}] \subset {\fra p}^{(m)}  \qquad 
[{\fra p}^{(m)}, {\fra p}^{(m)}] \subset {\fra h}^{(m)}   .
\ee
In the special case $\w_m=0$  the last equation reduces to $[{\fra p}^{(m)},
{\fra p}^{(m)}]=0$, and in this case ${\fra p}^{(m)}$ is not only a
subspace, but an an ideal. 

All algebras in the family ${\fra {so}}_{\w_1,\dots,\w_N}(N+1)$  (no matter
of how many values $\w_i$ are equal to zero) have an  asociated
Killing--Cartan metric form which can be defined in the same way as when the
algebra is simple, by the trace of the product of the adjoint representation
of the generators:
\be
g(\om_{ab},\om_{cd}) =
 \hbox{Trace} (\hbox{ad} \om_{ab} \cdot \hbox{ad}\om_{cd}).
\label{KillingCartanDef} 
\ee 
A simple calculation shows that the basis $\om_{ab}$ diagonalises the
Killing--Cartan form:
\be
g(\om_{ab},\om_{cd}) = - 2 (N-1) \delta_{ac} \delta_{bd} \w_{ab}.
\label{KillingCartan} 
\ee 

The Killing--Cartan form is only non-degenerate when the algebra is simple:
in fact, only when all $\w_i$ are different from zero the diagonal values
$g(\om_{ab},\om_{ab}) = - 2 (N-1) \w_{ab}$ are all different from zero. As
soon as a constant $\w_i$ is made zero, this property is lost. When a {\it
single} $\w_a=0$, then the subspace where the Killing--Cartan form vanishes
is exactly ${\fra p}^{(a)}$, while the restriction to the subalgebra
${\fra h}^{(a)}$ is non-degenerate. 

Each   ${\fra h}^{(m)}$ generates a subgroup $H^{(m)}$ of the CK group which
provides a quotient space 
\be 
{\cal S}^{(m)} \equiv  SO_{\w_1,\dots,\w_N}(N+1) \left/  \left( 
SO_{\w_1,\dots,\w_{m-1}}(m)\otimes
SO_{\w_{m+1},\dots,\w_N}(N+1-m) \right) \right. .
\label{dd}
\ee 
The dimension of ${\cal S}^{(m)}$ is that of ${\fra p}^{(m)}$; in fact,
${\fra p}^{(m)}$ is canonically identified with the tangent space to ${\cal
S}^{(m)}$ at the origin:
\be
 \mbox{dim}({\cal S}^{(m)})=m(N+1-m) .
\label{de}
\ee
Then ${\cal S}^{(m)}$ is  a symmetrical homogeneous space, and will be
generically  called  {\it orthogonal CK space}. There are exactly $N$ such
symmetrical homogeneous spaces ${\cal S}^{(1)}$,
${\cal S}^{(2)}$, \dots, ${\cal S}^{(N)}$ associated to each CK group
$SO_{\w_1,\dots,\w_N}(N+1)$; for the moment we will understand the values
$\w_1,\dots,\w_N$ as already fixed when we consider the aggregate of spaces
${\cal S}^{(m)}$. 

These $N$ spaces, although different, are not completely unrelated,  and it
is possible to reformulate all properties of any given space  in terms of any
other space, say ${\cal S}^{(m)}$, in the aggregate. Although this
possibility exists for any $m$, it is most easily understood when
interpreting ${\cal S}^{(2)},
\dots, {\cal S}^{(N)}$ in terms of ${\cal S}^{(1)}$.   As a matter of fact,
the spaces ${\cal S}^{(1)}$ associated to the CK algebras 
$\frak{so}_{\w_1,\dots,\w_N}(N+1)$ are rather well known (in particular when
the constants $\w_2, \w_3, \dots \w_N$ are all positive they  are the
constant curvature Riemannian spaces, cf.\ next Section). Now the key for
the  interpretation of ${\cal S}^{(2)}, \dots, {\cal S}^{(N)}$ in terms of
${\cal S}^{(1)}$ lies in the fact that the subgroups
$H^{(m)}$ ($m=1, 2, \dots, N$) are identified with the isotropy subgroups of
a point ($m=1$), a line ($m=2$),\dots, a hyperplane ($m=N$) in ${\cal
S}^{(1)}$.  Hence, ${\cal S}^{(1)}$ being taken as {\it the} space, its
elements being called {\it points}, ${\cal S}^{(2)}$ is the space of all
lines in 
${\cal S}^{(1)}$, ${\cal S}^{(3)}$ is the space of all 2-planes in 
 ${\cal S}^{(1)}$, and so on. This view allows a large freedom for
translating properties of any of the spaces ${\cal S}^{(2)}, {\cal S}^{(3)},
\dots, $  to properties of lines, 2-planes, \dots, in ${\cal S}^{(1)}$. In
some cases, this translation gives a much clearer picture than it would be
possible by blindly working with each space ${\cal S}^{(m)}$. 

An important feature of homogeneous spaces is their rank. 
 When dealing with homogeneous spaces associated to simple Lie groups, the
rank of the space (not to be confused with the rank of the group itself) is
usually defined as the maximal dimension of a totally geodesic flat
submanifold
\cite{Gilmore}. An alternative definition is preferable in the context of CK
homogeneous spaces: we will define the {\it rank} of a CK homogeneous space
${\cal S}^{(m)}$ as the number of independent invariants under the action of
the CK group for each generic pair of elements in the space
${\cal S}^{(m)}$. This number was first determined by Jordan \cite{Jordan}
when the group is the motion group of the
$N$-dimensional Euclidean space; it has a single invariant (the ordinary
distance) associated to each pair of points,   two invariants for each pair
of lines (an angle   and  a distance between the two lines),  and, in
general, $\mbox{min}( m,N+1-m)$ invariants for a pair of $(m-1)$-planes
(these invariants are stationary angles and a single stationary distance).
The reason why this definition of rank is better in the CK context is that
the total number of invariant stationary angles and distances obtained by
Jordan turns out to be the same for all spaces in the CK family, i.e., do not
depend on the values $\w_i$: 
\be
 \mbox{rank}({\cal S}^{(m)})=\mbox{min}( m,N+1-m).
\label{df}
\ee
 
In addition to the rank, which is an essential property making the spaces 
${\cal S}^{(1)}$, ${\cal S}^{(2)}$, \dots, ${\cal S}^{(N)}$ rather different
among themselves, the fact that they are homogeneous spaces of the same Lie
group entitles the space 
${\cal S}^{(m)}$ to inherit from its Lie algebra/group the following
geometrical structures
\cite{tesis}:
 
\noindent
$\bullet$ A  structure of symmetrical homogeneous space, which leads to a
canonical connection invariant under the CK group. 
 
\noindent
$\bullet$ A (possibly degenerate) quadratic main metric, coming from a
suitable rescaling of a Killing--Cartan form. The main metric in the space
${\cal S}^{(m)}$ is non-degenerate when the constants 
$\w_{1},\w_{2},\dots,
\w_{m -1},\w_{m +1},\dots, \w_{N}$ are all different from zero (note that
$\w_m$ is missing in this list); in this case its Levi--Civita connection
coincides with the canonical connection. 
 
\noindent
$\bullet$ When one of the constants 
 $\w_{1},\w_{2},\dots, \w_{m -1},\w_{m +1},\dots, \w_{N}$ is equal to zero,
then   the main metric is degenerate  and the space ${\cal S}^{(m)}$  has an
invariant foliation. This can be  considered  as a fibered structure, each
of whose leaves carries a subsidiary metric, coming again from the
Killing--Cartan form through restriction to the leaves and suitable
rescaling. 
  
\noindent
$\bullet$ Sectional curvatures of the space ${\cal S}^{(m)}$ are completely
determined by the value
$\w_{m}$. Only the rank-one spaces ${\cal S}^{(1)}$  and ${\cal S}^{(N)}$
are of {\it constant curvature}; CK spaces of higher rank are not of
constant curvature as this  is usually understood, yet their structure is as
close to constant curvature as a higher rank space can allow, because higher
rank spaces will have neccesarily to contain completely geodesic flat
submanifolds of dimension at least equal to the rank (and exactly equal to
the rank when the group is simple). 
 
\noindent
$\bullet$ Finally, the  canonical connection and the complete hierarchy of
subsidiary metrics are compatible.


\sect{Rank-one spaces}

For each CK group $SO_{\w_1,\dots,\w_N}(N+1)$, the CK space $\xxp$ is the
rank-one symmetrical homogeneous space obtained as the quotient by the
subgroup $H^{(1)}$: 
\be
\xxp\equiv SO_{\w_1,\dots,\w_N}(N+1) \left/ SO_{\w_2,\dots,\w_N}(N) \right. .
\ee
The dimension of the space $\xxp$ is $N$. The subgroup $H^{(1)}$ is
generated by the subalgebra ${\fra h}^{(1)}$ of the Cartan decomposition
associated to the involution $\Theta^{(1)}$:
${\fra {so}}_{\w_1,\dots,\w_N}(N+1)={\fra p}^{(1)}\oplus {\fra h}^{(1)}$.  
A good choice when dealing with a space of type ${\cal S}^{(m)}$ is to
replace the  notation $\om_{ab}$ by a  new `rank-adapted'  notation which
conveys the interpretation of generators either as translations or as
rotations in ${\cal S}^{(m)}$. Here, for $\xxp$ we denote the generators
$\om_{0i}$ in ${\fra p}^{(1)}$ as $P_i$ and those $\om_{ij}$ in
${\fra h}^{(1)}$ as
$J_{ij}$ $(i, j=1, \dots, N,\  i<j)$ according to the following arrangement:

\medskip

\noindent\hskip 0.5truecm
\begin{tabular}{|cccccc|ccccc}
$\om_{01} $&$ \om_{02} $& $ \om_{03} $& $ \ldots $& $ \om_{0N} $ 
&  &$P_1 $&$P_2 $& $P_3 $& $ \ldots $& $ P_N$\\
\cline{1-5}
\cline{7-11}
\multicolumn{1}{c}{\,}&$ \om_{12} $& $ \om_{13} $& $ \ldots $& $ \om_{1N} $ 
&\multicolumn{1}{c}{$\quad\equiv\quad$}&\multicolumn{1}{c}{\,}  
&$ J_{12} $& $ J_{13} $& $ \ldots $& $ J_{1N} $ \\
\multicolumn{1}{c}{\,}&$   $& $ \om_{23} $& $ \ldots $& $ \om_{2N} $ 
&\multicolumn{1}{c}{\,}&\multicolumn{1}{c}{\,} 
&$   $& $ J_{23} $& $ \ldots $& $ J_{2N} $ \\
\multicolumn{1}{c}{\,}&$   $& $   $& $ \ddots $& $ \vdots $ 
&\multicolumn{1}{c}{\,}&\multicolumn{1}{c}{\,} 
&$   $& $   $& $ \ddots $& $ \vdots $ \\
\multicolumn{1}{c}{\,}&$   $& $   $& $   $& $ \om_{N-1\,N} $ 
&\multicolumn{1}{c}{\,}&\multicolumn{1}{c}{\,} 
&$   $& $   $& $   $& $ J_{N-1\,N}$
\end{tabular}

\medskip\noindent 
and we call now $\k_i$ and $\k_{ij}$ the coefficients $\w_i$, and
$\w_{ij}$.  In this notation the non-zero Lie brackets of the CK algebra
(\ref{ca}) clearly display properties of translations and  rotations in
$\xxp$:
\be
\begin{array}{lll}
    [J_{ij},J_{ik}]=\k_{ij} J_{jk}   &\quad
        [J_{ij},J_{jk}]=-J_{ik}     &\quad 
        [J_{ik},J_{jk}]=\k_{jk}J_{ij}  \cr 
   [J_{ij},P_{i}]=P_j      &\quad 
       [J_{ij},P_{j}]=-\k_{ij}P_i  &\quad  \cr 
    [P_i,P_j]=\k_1\k_{1i} J_{ij}   &\quad  &\quad  
\end{array}
\ee
with $i<j<k$,\  $i,j,k=1,\dots,N$; a symbol like $\k_{11}$ with two equal
indices will be always understood as equal to 1. Notice that the constant
$\k_1$ {\it only} appears in the commutators of translations, foreshadowing
its role as the curvature of the space.

When the constant $\k_1$ is equal to zero, and only in this case, the CK
group acts on the space as a group of linear-affine transformations. In
other cases the group action is intrinsically non-linear. But it is possible
to {\it linearize} the action for {\it all} CK spaces $\xxp$ by going to
some ambient space. The tool to do this is the {\it vector}  representation
of $SO_{\w_1,\dots,\w_N}(N+1)$; in this representation the generators are
given by the matrices:
\be
\om_{0i} \to P_i^{(1)}=-\k_{0i}\,e_{0i}+e_{i0}   \qquad
\om_{ij} \to J_{ij}^{(1)}=-\k_{ij}\,e_{ij}+e_{ji} 
\ee
each of which satisfy the condition 
$X^T \Lambda_0^{(1)} + \Lambda_0^{(1)}  X$,
where $\Lambda_0^{(1)} $ is the matrix 
\be
\Lambda_0^{(1)} =\diag(1,\w_{01},\w_{02},\dots,\w_{0N}) .
\label{llaamm}
\ee
This representation of the Lie algebra produces the {\it vector}
representation of the CK group
$SO_{\w_1,\dots,\w_N}(N+1)$ as a group of matrices of order $N+1$, which acts
naturally and linearly (via matrix multiplication) in
$\R^{N+1}=(x^0,x^1,\dots,x^N)$. This action has two properties which are
relevant for our purposes:

\noindent
$\bullet$ It leaves invariant a quadratic form whose
matrix is $
\Lambda_0^{(1)}$

\noindent
$\bullet$ The subgroup $\hhp$  generated by the subalgebra ${\fra h}^{(1)}$
is the isotopy  subgroup of the  point $O=(1,0,\dots,0)\in\R^{N+1}$, i.e.,
the origin of  $\xxp$.

Hence, the space $\xxp$ can be identified with the orbit of $O$ under the
linear action of the group
$SO_{\w_1,\dots,\w_N}(N+1)$ in the  space $\R^{N+1}$. This action is by
isometries of the metric $\Lambda_0^{(1)}$, so it cannot be transitive in
$\R^{N+1}$, but only transitive on each orbit, which should be neccesarily
contained in the `sphere'
\be
(x^0)^2+\sum_{l=1}^N\k_{0l}(x^l)^2=1 .
\label{sphere}
\ee
The $N+1$ coordinates 
$(x^0,x^1,\dots,x^N)$ are called {\it Weierstrass coordinates}  for the CK
space
$\xxp$; their importance stems from the linear character of the group action
on them. There are two other natural coordinate systems in
$\xxp$, which are of a type called {\it geodesic} in differential geometry:

\noindent
$\bullet$ The point $\exp(a^1P_1)\exp(a^2P_2)\dots \exp(a^NP_N) O$ has
$(a^1,\dots,a^N)$ as {\it geo\-de\-sic parallel coordinates}.

\noindent
$\bullet$ The point $\exp(\te^N J_{N-1,N})\dots\exp(\te^2 J_{12})
\exp(\te^1P_{1})\,O$ has $(\te^1,\dots,\te^N)$ as   {\it geo\-de\-sic polar
coordinates}.

In particular, for the relationship between geodesic parallel and Weierstrass
coordinates we get
\be
x^0=\prod_{l=1}^N\! \Ck_{\k_{0l}}(a^l) \qquad
x^i= \Sk_{\k_{0i}}(a^i)\prod_{l=i+1}^N\! \Ck_{\k_{0l}}(a^l) \qquad
x^N=\Sk_{\k_{0N}}(a^N).
\label{paral}
\ee

When $\k_1=0$, the sphere (\ref{sphere})  reduces to an affine hyperplane in
$\R^{N+1}$ with equation
$x^0=1$, and geodesic parallel coordinates are simply cartesian coordinates
in this hyperplane. 

There are two ways to characterize the metric structure of $\xxp$. The
intrinsic one starts by  translating the Killing--Cartan form
(\ref{KillingCartan}) to  the rank-one language. The diagonal non-zero values
are:
$ g(P_i,P_i) = - 2 (N-1) \k_1 \k_{1i}, \  g(J_{jk},J_{jk}) = - 2 (N-1)
\k_{jk} $ so the restriction of the Killing--Cartan form to the subspace
${\fra p}^{(1)}$ can be written as:
\be
\left. g \, \right|_{{\fra p}^{(1)}} =- 2 (N-1) \k_1 g^{(1)}  
\ee
where $g^{(1)}$, the natural candidate for the metric in the tangent space
${\fra p}^{(1)}$, is:
\be
g^{(1)}(P_i,P_j) = \delta_{ij} \k_{1i} .
\ee  
The matrix of the metric $g^{(1)}$ in the tangent space
${\fra p}^{(1)}$ is, in the canonical basis $(P_1, P_2, \dots, P_N)$:
\be
\Lambda^{(1)}=\diag (+,\k_{12},\k_{13},\dots,\k_{1N})
=\diag (+,\, \k_2,\, \k_2\k_3,\dots,\, \k_2\!\cdots\!\k_N).
\label{mm}
\ee
Thus the constants $\k_2, \k_3, \dots, \k_N$ determine the signature of the
main metric at the origin in the space $\xxp$. This metric can be translated
to all points in the CK space
$\xxp$ by the group action, so that the action is by isometries. Now the
curvature of this metric can  be computed and turns out to be constant and
equal to $\k_1$.  This completes the geometric interpretation  of the
constants
$\k_i$ in the geometry of the space $\xxp$, and suggest a more detailed
notation for these rank-one spaces: $\xxp\equiv {\cal
S}^{[\k_1]\k_2,\dots,\k_N}$. 

When all the constants $\k_2, \k_3, \dots, \k_N$ are different from zero,
then the metric  is non-degenerate (Riemannian or  pseudoRiemannian case),
and it is definite positive when all of them are positive (Riemannian case).
Otherwise, when  a given
$\k_a=0$, $(a=2, \dots, N)$, the metric is degenerate and we introduce the
following  decomposition for ${\fra p}^{(1)} $: 
\be
{\fra p}^{(1)} = {\fra b_a}^{(1)} \oplus {\fra f}_a^{(1)} 
\qquad {\fra b}_a^{(1)}=\langle P_1,\dots,P_{a-1}\rangle\qquad
{\fra f}_a^{(1)}=\langle P_{a},\dots,P_{N}\rangle .
\ee 
In this case ${\fra f}_a^{(1)}$ is an ideal and the restriction of $g^{(1)}$
to this subalgebra vanishes.  Actually, whether or not $\k_a=0$, we always
have:
\be
\left. g^{(1)}  \, \right|_{{\fra f}_a^{(1)}} = \k_{1a} g_a^{(1)}  
\ee
where $g_a^{(1)}$ is defined in the subspace ${\fra f}_a^{(1)}$ as: 
\be
g_a^{(1)}(P_i,P_j) = \delta_{ij} \k_{ai} \qquad i,j=a, \dots, N 
\ee  
and this suggests to take $g_a^{(1)}$ as the metric in the subspace ${{\fra
f}_a^{(1)}}$ of the tangent space ${\fra p}^{(1)}$.  When $\k_a \neq 0$ no
real advantage is gained by considering
$g_a^{(1)}$ further to $g^{(1)}$, because they are simply proportional,  but
when $\k_a=0$, then the main metric vanishes when restricted to ${{\fra
f}_a^{(1)}}$ and the introduction of a {\it new} metric in this subspace is
meaningful.

These special properties of the subspace ${{\fra f}_a^{(1)}}$ of the tangent
space at the origin when $\k_a=0$ correspond to the existence of an invariant
foliation of the space
${\cal S}^{[\k_1]\k_2,\dots,\k_{a-1},0,\k_{a+1}\dots,\k_N}$ itself in this
case. This follows also quite clearly from the equation of the sphere
(\ref{sphere}), which reduces when
$\k_a=0$ to an equation involving only the variables $x^0, x^1, \dots,
x^{a-1}$. Each foliation leaf is coordinatised by the remaining variables
$x^a, x^{a+1},
\dots, x^N$, so that the subspace ${{\fra f}_a^{(1)}}$ is the  tangent space
at the origin to the foliation leaf through the origin. 

When there are more than one constant $\k_a$ equal to zero, we  have two
nested foliations, and the extension of the preceeding ideas to this case is
clear. 

So the picture emerging from this description is the following: the
qua\-dratic metric structure of the rank-one space
$\xxp$ is encoded in a main metric $g^{(1)}$ {\it and} a set of subsidiary
metrics, denoted $g_a^{(1)}$, one for each zero constant
$\k_a=0$ in the list $\k_2, \dots, \k_N$. These subsidiary metrics  are
defined in each leaf of the invariant foliation(s) associated to the zero
value of the constant(s) $\k_a=0$. 

In particular, when $\k_2=0$, the space ${\cal S}^{[\k_1]0,\k_3,\dots,\k_N}$
has an invariant foliation whose  set of leaves  is 
$(x^0)^2+\k_1 (x^1)^2=1 \equiv {\cal S}^{[\k_1]}$. Each leaf is described by
the set of all the values of $x^2,\dots,x^N$, and hence can be identified
with a CK space ${\cal S}^{[0]\k_3,\dots,\k_N}$. The main metric
$g^{(1)}$ is degenerate and  vanishes in each leaf. The subsidiary metric
$g_2^{(1)}$ is  well defined in each leaf. Should this look a bit involved,
the paradigmatic example of this case is the Galilean spacetime, discussed in
the last Section (cf.\ Table I below).

There is also an `extrinsic' way to compute the main metric in
$\xxp$: start from the CK group action in the ambient space
$\R^{N+1}$ as a group of isometries of the {\it flat} metric:
\be
(ds^2)_0^{(1)} = (dx^0)^2+\sum_{l=1}^{N}\k_{0l}(dx^l)^2  
\label{AmbientMetricRankOne}
\ee
whose relation with the Killing--Cartan form  is clear. As the space $\xxp$
is identified with the sphere (\ref{sphere}), it would be  obvious to
consider the restriction of this flat metric to the sphere. This restriction
turns out to be proportional to $\k_1$, and hence it vanishes when $\k_1=0$.
It is therefore natural to consider the restriction of
(\ref{AmbientMetricRankOne})  to the sphere taking out the factor
$\k_1$; this gives a well-defined non-trivial metric in all cases, no matter
on  whether
$\k_1$ is zero or not, and this metric coincides with the one derived earlier
by group theoretical reasoning. 

Once the main metric is known, it is a simple matter to translate  it to
other coordinates. We give such an expression for two different set of
coordinates. First, we define {\it Beltrami coordinates} for the CK space
$\xxp$ as:
\be
\eta^i := \frac{x^i}{x^0}\qquad i=1,\dots,N.
\ee
These coordinates (like Weierstrass ones) owe their name to the linear model
of hyperbolic space, where they were first introduced. The main metric is
given by:
\be
(ds^2)^{(1)} = 
\frac{(1+\k_1 \| \eta \| _\k^2 ) \, \| d\eta \| _\k^2 - \k_1 \langle \eta
| d\eta\rangle_\k^2 }{(1+\k_1 \|  \eta \| _\k^2  )^2 }
\label{zzaas}
\ee
with $\eta=(\eta^1,\dots,\eta^N)$, $d\eta=(d\eta^1,\dots,d\eta^N)$, and where
we have introduced two shorthands:
\be
\langle a | b \rangle_\k := a^1 b^1 + \sum_{i=2}^N \k_{1i} a^i b^i  \qquad 
\|  a \| _\k^2   := \langle a | a \rangle_\k .
\ee

Further to Beltrami coordinates, the next natural choice is the geodesic
parallel system of coordinates (\ref{paral}). Here the metric reads
\be
(ds^2)^{(1)} =\prod_{l=2}^N\!\Ck^2_{\k_{0l}}(a^l) \, (da^1)^2+
\sum_{i=2}^{N-1} \k_{1i}\!\prod_{l={i+1}}^N\!\!\Ck^2_{\k_{0l}}(a^l)\,
(da^i)^2+\k_{1N}\,(da^N)^2 . 
\label{metricparal}
\ee
When $\k_1=0$ but all other $\k_i \neq 0$ this reduces to the flat space
metric with suitable signature, $(ds^2)^{(1)}=(da^1)^2+
\sum_{i=2}^{N} \k_{1i} (da^i)^2$; in this case the space can be identified
with $\R^N$ and $a^i$ are cartesian coordinates.


\sect{Rank-two spaces}

We focus now on the CK spaces of the $\xxl$ type, which are  got by taking
the quotient of $SO_{\w_1,\dots,\w_N}(N+1)$ by the subgroup $H^{(2)}$:
\be 
\xxl \equiv  
SO_{\w_1,\dots,\w_N}(N+1)/(SO_{\w_1}(2)\otimes SO_{\w_3,\dots,\w_N}(N-1)).
\ee
The dimension of this space $\xxl$ is $2(N-1)$, and its rank is
$\mbox{min}(2,N-1)$, which is equal  to 2 except in the special case $N=2$,
where the rank is 1. In this Section we will understand we are  working in
the generic case $N>2$, so $\xxl$ will be actually a rank-two space. It is
symmetric since the subgroup
$H^{(2)}$ is generated by the subalgebra
${\fra h}^{(2)}$ of the Cartan decomposition provided by the automorphism
$\Theta^{(2)}$: ${\fra {so}}_{\w_1,\dots,\w_N}(N+1)={\fra p}^{(2)} \oplus
{\fra h}^{(2)}$ where
\be 
 {\fra p}^{(2)}=\langle \om_{0j},\ \om_{1j} \quad    j=2,\dots,
N\rangle,\quad
 {\fra h}^{(2)}= \langle \om_{01};\quad  \om_{ij}\ \ i,j=2,\dots,N \rangle .
\ee 
Hence we have in $\xxl$ two sets of $(N-1)$ translations and two types of
rotations. The structure of $\xxl$   can be more clearly appreciated  by
naming the abstract generators $\om_{ab}$ in the `rank-adapted' notation as
follows:
\medskip

\noindent 
\begin{tabular}{c|cccccc|cccc}
$\om_{01} $&$ \om_{02} $& $ \om_{03} $& $ \ldots $& $ \om_{0N} $ 
&  &$\!\! -J_{(1)(2)} $&$P_{(2)1} $& $P_{(2)2} $& $ \ldots $&  $
P_{(2)N-1}$\\
 &$ \om_{12} $& $ \om_{13} $& $ \ldots $& $ \om_{1N} $ 
&\multicolumn{1}{c}{$ \equiv $}& 
&$ P_{(1)1} $& $ P_{(1)2} $& $ \ldots $& $P_{(1)N-1} $ \\
\cline{2-5}
\cline{8-11}
\multicolumn{1}{c}{\,}&$   $& $ \om_{23} $& $ \ldots $& $ \om_{2N} $ 
&\multicolumn{1}{c}{\,}&\multicolumn{1}{c}{\,} 
&$   $& $ J_{12} $& $ \ldots $& $ J_{1\,N-1} $ \\
\multicolumn{1}{c}{\,}&$   $& $   $& $ \ddots $& $ \vdots $ 
&\multicolumn{1}{c}{\,}&\multicolumn{1}{c}{\,} 
&$   $& $   $& $ \ddots $& $ \vdots $ \\
\multicolumn{1}{c}{\,}&$   $& $   $& $   $& $ \om_{N-1\,N} $ 
&\multicolumn{1}{c}{\,}&\multicolumn{1}{c}{\,} 
&$   $& $   $& $   $& $ J_{N-2\,N-1}$
\end{tabular}

\medskip

Therefore we have introduced two  sets of indices $(a)$ and $i$ with ranges
$a=1,2$, and $i=1,\dots,N-1$. We complete the `rank-two' notation  by
denoting the  $\w_i$ coefficients as: 
\be
\w_1, \w_2, \w_3, \dots ,\w_N \quad \equiv \quad  
\k_{(2)}, \k_1, \k_2, \dots ,\k_{N-1}  
\ee
The value of these notational  changes is clear once we write the
commutation rules  (\ref{ca}) which now read:
\be
\begin{array}{ll}
[J_{ij},J_{ik}]  =\k_{ij}J_{jk}   & 
 [J_{ij},J_{jk}] =-J_{ik} \qquad\qquad  [J_{ik},J_{jk}] =\k_{jk}J_{ij}\cr
 [J_{ij},P_{(a)i}] =P_{(a)j}&[J_{ij},P_{(a)j}] =-\k_{ij}P_{(a)i}\cr
[J_{(1)(2)},P_{(1)i}] =P_{(2)i}&[J_{(1)(2)},P_{(2)i}] =-\k_{(2)}P_{(1)i}\cr
[P_{(1)i},P_{(1)j}] =\k_1\k_{1i} J_{ij} &
[P_{(2)i},P_{(2)j}] =\k_1\k_{1i}\k_{(2)}J_{ij} \cr
[P_{(1)i},P_{(2)i}]  =\k_1\k_{1i}\k_{(2)} J_{(1)(2)}. & 
\end{array}
\label{CommRelsRankTwo}
\ee
Here again any two-index coefficient with two equal indices (as $\k_{11}$)
will be always assumed as equal to 1. Note that the constant $\k_1$ (the old
$\w_2$) is now appearing in all the commutators of the rank-two translations
$P_{(a)i}$, foreshadowing again its role as the curvature of the space
$\xxl$. 

An important step in the rank-one  case is the introduction of an ambient
space on which the CK group acts linearly, and where $\xxp$ is embedded (the
same idea can be succesfully applied for all CK spaces of any rank). All we
need to carry out this idea is to replace the vector representation of the
CK  group (\ref{cd}) by another representation, acting linearly on  some
space and having the subgroup $H^{(2)}$ as the isotropy subgroup. This is
accomplished by taking the antisymmetrized  square of the vector
representation, which for the brevity sake will be called here the {\it
bivector} representation of the CK group. Consider first the following
$\frac 12 N(N+1)\times \frac 12 N(N+1)$ matrices where the matrix  indices
are pairs $ij$ ($i<j$) of indices in the set $0, 1, \dots, N$:
\be
e_{ij,kl} \qquad i<j \qquad k<l \qquad i,j,k,l=0,1,\dots ,N 
\ee
with an entry $1$ in  the row $ij$ and column $kl$, with 0's otherwise. The
explicit form of the  bivector representation  of the CK algebra,
distinguised by a $^{(2)}$ superscript is:
\bea
 &&\!\!\!\!\!\!\!
J_{ij}^{(2)}=-\k_{ij}\sum_{s=j+2}^N \! e_{i+1s,j+1s}+\sum_{s=j+2}^N\!
e_{j+1s,i+1s} 
 -\k_{ij}\sum_{s=0}^i e_{si+1,sj+1}+\sum_{s=0}^i e_{sj+1,si+1} \cr
 &&\!\!\!\!\!\!\!
\qquad\qquad +\k_{ij}\sum_{s=i+2}^j \!e_{ i+1s,s j+1 }-
\sum_{s=i+2}^j\! e_{sj+1,i+1s}\cr 
 &&\!\!\!\!\!\!\!
 J_{(1)(2)}^{(2)}=\k_{(2)}\sum_{s=2}^Ne_{0s,1s}-\sum_{s=2}^N
e_{1s,0s}\label{GenBivectorRepres}
\\
 &&\!\!\!\!\!\!\!
 P_{(1)j}^{(2)}=-\k_{0j}\sum_{s=j+2}^N\!
 e_{1s,j+1s}+\sum_{s=j+2}^N\! e_{j+1s,1s} 
  -\k_{0j} \, e_{01,0j+1}+  e_{0j+1,01} \cr
 &&\!\!\!\!\!\!\!
 \qquad\qquad +\k_{0j}\sum_{s=2}^j e_{  1s,s j+1 }-\sum_{s= 2}^j e_{sj+1, 
1s}\cr  
 &&\!\!\!\!\!\!\!
P_{(2)j}^{(2)}=-\k_{(2)}\k_{0j}\!\!\sum_{s=j+2}^N\!\! e_{0s,j+1s}+
\!\!\sum_{s=j+2}^N\!\! e_{j+1s,0s}
+\k_{(2)}\k_{0j}\sum_{s=1}^j e_{ 0s,s j+1 }-\sum_{s= 1}^j e_{sj+1,  0s}  .
\nonumber
\eea

In the previous Section, the vector representation of the CK algebra
generated a group of linear isometries in the ambient space
$\R^{N+1}$, relative to the metric matrix
$\Lambda^{(1)}_0$ (\ref{llaamm}) and whose isotropy subgroup was $H^{(1)}$.
For the bivector representation we have similar properties, as well as a new
one. First, each rank-two generator $X$ (\ref{GenBivectorRepres}) satisfies
the condition $X^T \Lambda_0^{(2)} + \Lambda_0^{(2)} X$, where
$\Lambda_0^{(2)}$ is the $\frac 12 N(N+1)\times \frac 12 N(N+1)$ matrix 
\bea
 &&\Lambda_0^{(2)} =e_{01,01}+\sum_{i=1}^{N-1}\k_{0i}\,e_{0i+1,0i+1}
+\k_{(2)}\sum_{i=1}^{N-1}\k_{0i}\,e_{1i+1,1i+1} \cr
&&\qquad + \k_{(2)}{\sum_{i,j=1; i<j}^{N-1}}
\k_{0i}\k_{0j}\,e_{i+1j+1,i+1j+1} .
\label{bivectorAmbEspMetric}
\eea

By direct exponentiation of the generators (\ref{GenBivectorRepres})  we get
a group of $\frac 12 N(N+1)\times
\frac 12 N(N+1)$ matrices, the bivector representation of
$SO_{\w_1,\dots,\w_N}(N+1)$. This group acts linearly (by matrix
multiplication) in the auxiliar `bivector' ambient space
$\R^{N(N+1)/2}=(x^{ij})$ $(i<j,\ i,j=0,\dots N)$, as isometries  of the
metric (\ref{bivectorAmbEspMetric}), and therefore the action is not
transitive. It is also clear that the subgroup $H^{(2)}$ is the isotropy
subgroup of the point $O$ with $x^{01}=1$ and all other coordinates
$x^{ij}=0$ which will be taken as the origin. As a first step to the
determination of orbits, we first consider the sphere in $\R^{N(N+1)/2}$
corresponding to the metric (\ref{bivectorAmbEspMetric}):
\be
  (x^{01})^2+\sum_{i=1}^{N-1}\!\k_{0i}\,(x^{0i+1})^2  
+\k_{(2)}\!\sum_{i=1}^{N-1}\!\k_{0i}\, (x^{1i+1})^2   
 + \k_{(2)} \!\!  {\sum_{i,j=1;i<j}^{N-1}}\!\!
\k_{0i}\k_{0j}\, (x^{i+1j+1})^2=1 .
\label{rank2sphere}
\ee
The coordinates $x^{ij}$ in the ambient space are called  {\it Pl\"ucker
coordinates} for the space $\xxl$; they are the rank-two version of
Weierstrass coordinates. All this seems quite analogous to the  rank-one
case. However there is an essential difference. The space $\xxl$ is to be
identified with the orbit of $O$ under the group action; this orbit, with
dimension $2(N-1)$ cannot fill the sphere, of  dimension $N(N+1)/2-1$. This
is due to a new fact for the bivector representation of the CK group, which
leaves invariant not only the quadratic form  (\ref{bivectorAmbEspMetric})
but also a set of quadratic relations known as {\it Pl\"ucker relations}
(also Grassmann relations or even $p$-relations):
\be
x^{ij}x^{kl}-x^{ik}x^{jl}+x^{il}x^{jk}=0 \qquad i<j<k<l  \qquad
i,j,k,l=0,\dots,N 
\label{zb}
\ee
so actually the rank-two CK  space $\xxl$ should be identified to the
intersection of the sphere (\ref{rank2sphere}) with the family of quadratic
cones (\ref{zb}) in the bivector ambient space. We can profit from this new
fact: in some open neighbourhood of the origin of $\xxl$, we can take 
$x^{0j}$ and $x^{1j}$ ($j=2,\dots,N$) as the $2(N-1)$  independent
coordinates of the space. The coordinate  $x^{01}$ will be left as a
non-independent one, and the remaining will be eliminated by using the
Pl\"ucker equations with indices $01kl$:
\be
x^{kl}=\frac{x^{0k}x^{1l}-x^{0l}x^{1k}}{x^{01}} \qquad k,l=2,\dots,N.
\label{zc}
\ee
With this choice for the inessential coordinates $x^{kl},  (k,l=2,\dots,N)$
it can be indeed shown that {\it all} the Pl\"ucker equations (and not only
those with indices $01kl$) become identities. After this is done, everything
is similar to the rank-one case: we have some essential coordinates $x^{0i},
x^{1i}, \ i=2,
\dots, N-1$ and a single auxiliar inessential coordinate  $x^{01}$ which can
be eliminated by the equation of the sphere after (\ref{zc}) has been used. 

Once we have got an explicit description of the space $\xxl$, we turn to its
quadratic metric. Similarly to the rank-one case, there is a main and an
eventual set of subsidiary metrics in the space
$\xxl$, all of which are required in order to give a complete description of
its metric structure. The main metric $g^{(2)}$ comes again from the
Killing--Cartan form on the algebra (\ref{KillingCartan}), when suitably
restricted to the tangent space ${\fra p}^{(2)}$ to the  origin of $\xxl$. 
We first give the expressions for the diagonal non-identically zero values
of the Killing--Cartan form 
\be
\begin{array}{ll}
 g(P_{(1)i},P_{(1)i}) = - 2 (N-1) \k_1 \k_{1i}  &\quad 
g(P_{(2)i},P_{(2)i}) = - 2 (N-1) \k_{(2)} \k_1 \k_{1i} \cr
  g(J_{(1)(2)},J_{(1)(2)}) = - 2 (N-1) \k_{(2)} &\quad 
g(J_{jk},J_{jk}) = - 2 (N-1) \k_{jk} 
\end{array}
\label{KillingCartanRankTwo} 
\ee
where the constant $\k_1$ appears again only in the  restriction to the
${\fra p}^{(2)}$ subspace. It is clear that the natural candidate for the
metric in the tangent space ${\fra p}^{(2)}$ is obtained by writing the
restriction of the Killing--Cartan in the Lie algebra to the subspace
${{\fra p}^{(2)}}$ as:
\be
\left. g \, \right|_{{\fra p}^{(2)}} =  - 2 (N-1)\k_1 g^{(2)}
\ee
that is:
\be
g^{(2)}(P_{(1)i},P_{(1)i}) = \k_{1i}   \qquad 
g^{(2)}(P_{(2)i},P_{(2)i}) = \k_{(2)} \k_{1i} .
\ee  
At the origin of the space $\xxl$, and in the basis of the tangent space
provided by the translation generators $P_{(a)i}$ themselves,  the main
metric
$g^{(2)}$ is given by the matrix:
\be
\Lambda^{(2)}= 
\left(\begin{array}{cc}
\Pi&0 \\ 0&\k_{(2)}\Pi  \end{array}\right) \qquad
\Pi=\diag (+,\,\k_{12},\,\k_{13},\,\dots,\, \k_{1\,N-1})
\label{Lambda2}
\ee
so the signature is  determined by the constants $\k_{(2)}, \k_2, \dots,
\k_{N-1}$. This metric can be translated to all points in the CK space
 $\xxl$ by the group action, so that the CK group acts by isometries in
$\xxl$. Now the sectional curvatures of this metric can be computed, and the
result can be foreseen from the commutation relations
(\ref{CommRelsRankTwo}): sectional curvatures are not constant, but they are
as close to constant as a rank-two space would allow. At the origin, the
sectional curvature of the space
$\xxl$ along any 2-plane direction spanned by any two tangent vectors
 $P_{(a)i}, P_{(a)j}$ or $P_{(a)i}, P_{(b)i}$ is constant  and equal to
$\k_1$ (two vectors where both indices are different span a 2-plane for
which the curvature is always identically equal to zero, no matter of the
values of the constant $\k_1$; this is behind the classical definition of
the rank of the homogeneous space associated to a simple group).  This
completes the geometric interpretation of the constants $\k_i$ in the
geometry of the space
$\xxl$, and suggest a complete notation for these rank-two spaces:
$\xxl\equiv {\cal S}^{\k_{(2)}[\k_1]\k_2\dots\k_{N-1}}$. 

From this point onwards things are rather similar  to the rank-one case.
Invariant foliations appear when any $\k$ in the set $\k_{(2)}, \k_2, \k_3,
\dots, \k_{N-1}$ is equal to zero (when the main metric is degenerate). We
start by introducing a decomposition of ${\fra p}^{(2)}$ as: 
\be
\begin{array}{ll}
{\fra p}^{(2)} = {\fra b_a}^{(2)} \oplus {\fra f}_a^{(2)}  &\quad 
a=(2), 2, \dots N-1\cr
{\fra b}_{(2)}^{(2)}=\langle P_{(1)1},\dots,P_{(1)N-1}\rangle
&\quad 
{\fra b}_{a}^{(2)}=\langle
P_{(1)1},\dots,P_{(1)a-1}\,;\, P_{(2)1},\dots,P_{(2)a-1}\rangle\cr
{\fra f}_{(2)}^{(2)}=\langle P_{(2)1},\dots,P_{(2)N-1}\rangle
&\quad 
{\fra f}_{a}^{(2)}=\langle
P_{(1)a},\dots,P_{(1)N-1}\,;\, P_{(2)a},\dots,P_{(2)N-1}\rangle .
\end{array}
\ee
 When $\k_{(2)}=0$, (resp.\ $\k_a=0$) then
${\fra f}_{(2)}^{(2)}$ (resp.\ ${\fra f}_a^{(2)}$) is an ideal 
and the restriction of $g^{(2)}$ to this subalgebra vanishes. 
Whether or not $\k_{(2)}=0$ or $\k_a=0$, we always have:
\be
\left. g^{(2)}  \, \right|_{{\fra f}_{(2)}^{(2)}} = \k_{(2)} g_{(2)}^{(2)} 
\qquad   
\left. g^{(2)}  \, \right|_{{\fra f}_a^{(2)}} = \k_{1a} g_a^{(2)}   \qquad
a=2,
\dots, N-1
\label{metricRedRankTwo1}
\ee
where $g_{(2)}^{(2)}$ is defined in ${\fra f}_{(2)}^{(2)}$ (resp. 
$g_a^{(2)}$ is defined in ${\fra f}_a^{(2)}$) as: 
\be
\begin{array}{l}
  g_{(2)}^{(2)}(P_{(2)i},P_{(2)j}) = \delta_{ij} \k_{1i}  \qquad 
i,j=1, \dots, N-1 \cr
 g_a^{(2)}(P_{(1)i},P_{(1)j}) = \delta_{ij} \k_{ai}  \quad 
  g_a^{(2)}(P_{(2)i},P_{(2)j}) = \delta_{ij} \k_{(2)} \k_{ai}  \quad 
i,j=a, \dots, N\!-\!1. 
\end{array}
\label{metricRedRankTwo2}
\ee
Even when $\k_{(2)}=0$ or $\k_a=0$, these define metrics in the subspaces
${\fra f}_{(2)}^{(2)}$   or ${\fra f}_a^{(2)}$. In this case the complete
metric description of the space splits into a degenerate main metric, and a
metric in each of the fibers.  When a constant $\k_{(2)}, \k_2, \dots,
\k_{N-1}$ is zero, the ad-invariance of the corresponding subalgebra ${\fra
f}_{(2)}^{(2)}$ or ${\fra f}_a^{(2)}$ produce invariant foliations in the
rank-two space. For instance, when  $\k_{(2)}=0$, the equation
(\ref{rank2sphere}) is:
\be
(x^{01})^2+\k_{01}\,(x^{02})^2+ \dots +\k_{0N-1}\,(x^{0N})^2=1 
\ee
so the set of leaves (the base space for the fibered structure)  has
dimension $N-1$ and can be identified with a rank-one space ${\cal
S}^{[\k_1]\k_2,\dots,\k_{N-1}}$.  Each leaf in the foliation (the fiber),
characterized by some fixed values of the coordinates $x^{02},\dots,x^{0N}$,
is described by the remaining essential coordinates
$x^{12},x^{13},\dots,x^{1N}$;   the fiber can be identified with the rank-one
space ${\cal S}^{[0]\k_2,\dots,\k_{N-1}}$. 

When $\k_{2}=0$, the equation (\ref{rank2sphere}) is:
\be
(x^{01})^2+\k_{1}\,(x^{02})^2+ \k_{(2)}\k_{1}\,(x^{12})^2=1
\ee
so the set of leaves has now dimension $2$, and can be identified with  the 
two-rank space  ${\cal S}^{\k_{(2)}[\k_1]}$. A set of coordinates for each
leaf is specified  by the remaining  $x^{03},x^{04},\dots,x^{0N}$ and
$x^{13},x^{14},\dots,x^{1N}$. As a CK space the fiber is the rank-two space
is ${\cal S}^{\k_{(2)}[\k_1]\k_3,\dots,\k_{N-1}}$. The situation is similar
when $\k_3=0, \dots$, etc.

An alternative  approach to compute these metrics is to start from the flat
metric   in the bivector ambient space, then restrict it to the intersection
of the sphere (\ref{rank2sphere}) with the Pl\"ucker cones and take out the
coefficient of $\k_1$ in the expression thus obtained. The flat bivector
ambient metric is: 
\bea 
&& (ds)_0^{(2)}  = 
(dx^{01})^2 + \k_1 \sum_{i=1}^{N-1}\! {\k_{1i}}\,(dx^{0i+1})^2 +
    \k_1 {\k_{(2)}} \sum_{i=1}^{N-1}\! {\k_{1i}} \, (dx^{1i+1})^2 \cr   
&&\qquad\qquad +  \k_1 {\k_{(2)}}  {\sum_{i,j=1;i<j}^{N-1}}\!\!
 {\k_{1i}\k_{0j}} \, (dx^{i+1j+1})^2  .
\label{zg}
\eea
Working locally in some open neighbourhood of the origin in
$\xxl$ (determined by the condition $x^{01}>0$), we can expressed (\ref{zg}) 
in terms of $x^{01}, x^{0i}, x^{1i}$ by means of (\ref{zc});  then we
restrict to the sphere. We carry out this programm for the so-called {\it
Beltrami coordinates} in the rank-two space  defined by 
\be
\eta^i:=\frac{x^{0i+1}}{x^{01}} \qquad \xi^i:=\frac{x^{1i+1}}{x^{01}} 
\qquad i=1,\dots ,N-1.
\ee
The equation of the sphere (\ref{rank2sphere}) turns into
\be
(x^{01})^2 \, ({1+\k_1\| \> (\eta,\>\xi) \| _\k^2}) =1
\ee
where we introduce a notational shorthand, analogous to those introduced in
the rank-one case: for 
$\>\eta=(\eta^1,\dots,\eta^{N-1})$, 
$\>\xi=(\xi^1,\dots,\xi^{N-1})$, $\| ( .,.) \| _\k$ is defined by
\be
\| (\>\eta,\>\xi)\| _\k^2:=\sum_{i=1}^{N-1} {\k_{1i}} \,(\eta^i)^2  
+  {\k_{(2)}} 
\sum_{i=1}^{N-1}\! {\k_{1i}}\, (\xi^i)^2   
+ {\k_{(2)}} \!\! 
 {\sum_{i,j=1;i<j}^{N-1}}\!\!\!
 {\k_{1i}\k_{0j}} \, (\eta^i\xi^j-\eta^j\xi^i)^2 .
\ee
We remark that the expression $\| (\>\eta,\>\xi) \| _\k$ is only a  norm in
the standard sense of the term (i.e., definite positive)  when all $\k$
constants involved $\k_{(2)}, \k_2, \dots,  \k_{N-1}$ are positive.
Otherwise it is either indefinite or degenerate.  We also introduce the
companion notational shorthand:
\bea
&& \langle(\eta,\xi)|(d\eta, d\xi)\rangle_\k  :=
\sum_{i=1}^{N-1} {\k_{1i}} \,\eta^i \,d\eta^i   
+\k_{(2)}\sum_{i=1}^{N-1} {\k_{1i}} \, \xi^i\, d\xi^i   \cr 
&&\qquad\qquad\qquad\quad + \k_{(2)} {\sum_{i,j=1;i<j}^{N-1}}
 {\k_{1i}\k_{0j}} \,(\eta^i\xi^j-\eta^j\xi^i)
\,  d(\eta^i\xi^j-\eta^j\xi^i) . 
\eea
with $d\>\eta=(d\eta^1,\dots,d\eta^{N-1})$ and
$d\>\xi=(d\xi^1,\dots,d\xi^{N-1})$, 
as well as 
\be
 \| (d\eta, d\xi) \| _\k^2 :=
\sum_{i=1}^{N-1}\! {\k_{1i}} (d\eta^i)^2  
+\k_{(2)}\!\sum_{i=1}^{N-1}\! {\k_{1i}}   (d\xi^i)^2   
  + \k_{(2)}\!\!\! {\sum_{i,j=1; i<j}^{N-1}}\!\!\!
 {\k_{1i}\k_{0j}} \bigl( d(\eta^i\xi^j-\eta^j\xi^i)\bigr)^2 .
\ee
Then we find that the flat ambient metric, when restricted to $\xxl$ is
proportional to $\k_1$; by taking out this factor the end result for  the
main metric (\ref{zg}) in the space
$\xxl$ is expressed as
\be
(ds^2)^{(2)} =
\frac { (1+\k_1\| (\eta,\xi) \| _\k^2)\, \| (d\eta,d\xi) \| _\k^2 -\k_1\,
\langle (\eta,\xi)|(d\eta,d\xi)\rangle_\k^2}
{(1+\k_1\| (\eta,\xi) \| _\k^2)^2}.
\label{zh}
\ee
Notice the close ressemblance of this metric with its rank-one analogous
(\ref{zzaas}). 

A last comment is in order. In general, the Pl\"ucker relations
$x^{ij}x^{kl}-x^{ik}x^{jl}+x^{il}x^{jk}=0$ are invariant under the bivector
representation, but the r.h.s. of the relation is not. An exception is the
lowest dimensional rank-two case $N=3$. Here there is a single Pl\"ucker
relation $0123$, and the quadratic form
$x^{01}x^{23}-x^{02}x^{13}+x^{03}x^{12}$ is also invariant;  this is related
to the (exceptional) known quadratic Riemannian metric in the real
Grassmannian of two-planes in four dimensions. In any other real
Grassmannian, there is up to a factor a unique Riemannian quadratic metric,
which should coincide with the one we have derived.


\section{Curvature and metric in spacetimes and phase spaces}

We now turn to the physical interpretation for some of the  rank-two spaces
we have studied here: just as the rank-one CK spaces afford models for all
homogeneous spacetimes (e.g., those in the Bacry and Levy--Leblond
classification \cite{BLL}), their corresponding rank-two spaces gives models
for the phase spaces of a  free system whose spacetime is a rank-one CK
space. We will insist on the metric aspect, as usually the metric structure
of the phase space is disregarded in favour of the symplectic  structure,
upon which we have said nothing, and the metric structure which naturally
appear in our scheme cannot be easily guessed from what is more or less
implicit in the literature when dealing with phase spaces for curved
spacetimes. We restrict here to pointing out the most relevant traits. 

First, a rather elementary remark is that the grouping of the $2(N-1)$
coordinates of a rank-two space into two sets is intimately related to the
existence of pairs of canonically conjugated variables: the {\it
momentum}-like $\eta^i$ and {\it position}-like  $\xi^i$  coordinates.
Second, a warning: the notation we have tailored for rank-two spaces tries
to convey the meaning of quantities in the most close form as possible to
the rank-one case. We have used in both cases the name $\k_1$ for the
curvature of the space. This is satisfactory as long as we deal with either
rank-one {\it or} rank-two spaces alone, but turns rather confusing when we
insist on considering simultaneously associated rank-one and rank-two
spaces. In this case it is far better to go back to the `neutral' $\w$
notation used in the Section 2 of the paper. Each constant $\w_i$ will be
interpreted differently according as we are working in
$\xxp$ or $\xxl$; the complete notation for these spaces, 
$\xxp \equiv {\cal S}^{[\w_1]\w_2,\w_3,\dots, \w_N }$ or $\xxl \equiv {\cal
S}^{\w_1[\w_2]\w_3,\dots, \w_N }$ is informative enough to clear any
misunderstanding. 

We display  in Table I nine especially relevant CK spaces $\xxp$ in the
general  $N$-dimensional case. These are the spaces 
${\cal S}^{[\w_1] \w_2, +, \dots, +}$ associated to the algebras 
$\frak{so}_{\w_1, \w_2,+, \dots,+}(N+1)$ where the constants $\w_3, \dots,
\w_N$ are all positive. The first row gives  the usual  Riemannian spaces
with a non-degenerate and positive definite metric. The six remaining spaces
are the six well known possible homogeneous kinematical spacetimes in
$(N-1)+1$ dimensions. In the second row we find the  `absolute-time' models
associated to Newtonian spacetimes with the three possible values of the
spacetime curvature (oscillating Newton--Hooke (NH), with positive spacetime
curvature and group $T_{2N-2}(SO(N-1)\otimes SO(2))$; Galilei, with zero
curvature; and expanding NH, with negative curvature and group
$T_{2N-2}(SO(N-1)\otimes SO(1,1))$). These three spaces have a main
degenerate metric (whose length is the absolute time) and  a subsidiary well
defined metric (the purely spatial metric, which only  makes sense when taken
on each of the leaves of the invariant foliation, here  the leaves of
absolute simultaneity). In the third row the `relative-time' spacetime models
corresponding to relativistic spacetimes with a Lorentz type metric appear.

\medskip\medskip

{\footnotesize

 \noindent
{{\bf Table I.} The $N$-dimensional rank-one spaces ${\cal
S}^{[\w_1]\w_2,+,\dots, + }$.}
\smallskip

\noindent 
\begin{tabular}{lll}
\hline
\hline
\multicolumn{3}{l}{\qquad\qquad Riemannian spaces  ${\cal S}^{[\w_1]+,\dots,
+ }$:\quad  $\Lambda^{(1)}=\diag (+,+,\dots,+)$} \\
\multicolumn{3}{l}{\qquad\qquad No invariant foliation} \\
Elliptic Space&Euclidean Space&Hyperbolic Space\\
${\cal S}^{[+]+,\dots, +}  \simeq {\bf S}^N$&
${\cal S}^{[0]+,\dots, +}  \simeq {\bf  E}^N$&
${\cal S}^{[-]+,\dots, +}   \simeq {\bf  H}^N$\\
$SO(N+1)/SO(N)$&$ISO(N)/SO(N)$&$SO(N,1)/SO(N) $\\
Positive Curvature & 
Zero Curvature & 
Negative Curvature \\ 
No Invariant Foliation &
No Invariant Foliation &
No Invariant Foliation \\
\hline
\multicolumn{3}{l}{SemiRiemannian spaces ${\cal S}^{[\w_1]0,+,\dots,+}$:\   
$\Lambda^{(1)}=\diag (+,0,\dots,0)$ \ 
 $\Lambda_2^{(1)} = \diag (+,\dots,+)$ }\\
\multicolumn{3}{l}{Invariant foliation:\quad 
Set of foliation leaves ${\cal S}^{[\w_1]}$  \quad 
Fiber ${\cal S}^{[0],+,\dots,+}$} \\  
Oscillating NH Spacetime&Galilean Spacetime&Expanding NH Spacetime\\
${\cal S}^{[+]0,+,\dots, +}  $&
${\cal S}^{[0]0,+,\dots, +}  $&
${\cal S}^{[-]0,+,\dots, +}  $\\
$ONH/ISO(N\!-\!1)  \!\!$&
$IISO(N\!-\!1)/ISO(N\!-\!1)  \!\!$&
$ENH/ISO(N\!-\!1)  \!\!$\\
Positive Curvature & 
Zero Curvature & 
Negative Curvature \\ 
Invariant Foliation &
Invariant Foliation &
Invariant Foliation \\
Base ${\cal S}^{[+]}$, Fiber ${\cal S}^{[0],+,\dots,+}$ & 
Base ${\cal S}^{[0]}$, Fiber ${\cal S}^{[0],+,\dots,+}$ & 
Base ${\cal S}^{[-]}$, Fiber ${\cal S}^{[0],+,\dots,+}$ \\
\hline
\multicolumn{3}{l}{\qquad\qquad PseudoRiemannian spaces ${\cal
S}^{[\w_1]-,+,\dots, + }$:\quad  $\Lambda^{(1)}=\diag (+,-,\dots,-)$} \\
\multicolumn{3}{l}{\qquad\qquad No invariant foliation} \\
Anti-DeSitter Spacetime&Minkowskian Spacetime&DeSitter Spacetime\\
${\cal S}^{[+]-,+,\dots, +}  $&
${\cal S}^{[0]-,+,\dots, +}  $&
${\cal S}^{[-]-,+,\dots, +}  $\\
$SO(N\!-\!1,2)/SO(N\!-\!1,1) \!\!\!\!$&
$ISO(N\!-\!1,1)/SO(N\!-\!1,1) \!\!\!\!$&
$SO(N,1)/SO(N\!-\!1,1) \!\!\!\!$ 
\\ Positive Curvature & 
Zero Curvature & 
Negative Curvature \\ 
No Invariant Foliation &
No Invariant Foliation &
No Invariant Foliation \\
\hline
\hline
\end{tabular}}

\medskip
\medskip

In terms of the rank-one spaces, ${\cal S}^{[\w_1]\w_2,+,\dots,+}$,  $\w_1$
is the curvature and $\w_2$ determines the signature of the main metric.  A
non-zero positive
$\w_1$ is related to the oscillating NH or Anti-DeSitter radius as
$\w_1=1/R^2$ or to the expanding NH or DeSitter characteristic  length when
it is negative,  $\w_1=-1/R^2$.  The main metric is definite positive when
$\w_2>0$ --i.e., in the Euclidean, elliptic, and hyperbolic spaces--,
degenerate when
$\w_2=0$ --i.e., in the Galilean and both NH spacetimes-- and  Lorentzian
type when
$\w_2<0$ --i.e., in Minkowski and both DeSitter spacetimes--;  the relation
of
$\w_2$ with the standard relativistic constant $c$ is $\w_2=-1/c^2$.

In the kinematical spaces, and in parallel coordinates, $a^1$ should be
identified with the time coordinate, the others being space coordinates. In
particular, in the three $(1+3)$ Newtonian spacetimes ($N=4,
\k_2=0, \k_3=1, \k_4=1$), the main metric (\ref{metricparal}) reduces to
$(ds^2)^{(1)} =(da^1)^2$, which gives a length which depends only on the end
points of the curve, but not on the path itself. The leaves of the foliation
are characterized by
$a^1$ constant, and should be identified in the three cases with  an
Euclidean three-space, the  subsdiary metric being given by 
$g_2^{(1)}=\frac{1}{\k_2} g^{(1)} |_{{\fra f}_2^{(1)}}$:
\be
(ds^2)_2^{(1)}=  (da^2)^2+ (da^3)^2+ (da^4)^2 \qquad (a^1\ \mbox{constant}).
\ee

We now describe the metric structure of the phase spaces associated to the
nine CK rank-one spaces 
${\cal S}^{[\w_1]\w_2,+,\dots, + }$ of Table I whose corresponding rank-two
spaces (identified with the space of lines in ${\cal
S}^{[\w_1]\w_2,+\dots,+}$) are ${\cal S}^{\w_1[\w_2]+,\dots,+}$. Their
geometric properties, as far as their metrics are concerned, depend also on
the values of the constants $\w_1,
\w_2$ but in a rather different way to which it was formerly the case. Here
the constant $\w_1$ appears in the signature of the main metric, so that
whenever $\w_1=0$ the main metric is degenerate, (and we are in presence of
an invariant foliation in the phase space). This happens for the three
rank-two spaces
${\cal S}^{0[\w_2]+,\dots,+}$ with $\w_1=0$; depending on whether $\w_2>0,
\w_2=0, \w_2=-1/c^2<0$, these spaces are the sets of lines in the Euclidean,
Galilean and Minkowskian rank-one spaces. All these phase spaces ${\cal
S}^{0[\w_2]+,\dots,+}$ have an invariant foliation, with an
$(N-1)$-dimensional base space and
$(N-1)$-dimensional leaves. The base space is the rank-one space ${\cal
S}^{[\w_2]+,\dots,+}$ identified in kinematical terms with the  {\it
velocity} space, with momentum-like coordinates
$\eta^{1},\dots,\eta^{N-1}$. The degenerate main metric (\ref{zh}) reduces
 to a metric in the base $(N-1)$-velocity space.  Each leaf in the foliation
is coordinatised by the values of the position-like coordinates
$\xi^{1},\dots,\xi^{N-1}$. All these results should have been expected: when
$\w_1=0$ the rank-one spacetime has zero curvature; in this case parallelism
 of lines is absolute, and we can meaningfully class all lines into
parallelism classes, each of which is completely described by its (common)
velocity.  In particular, when $\w_1=0$, $\w_3=\w_4=1$ and $N=4$ the
expression for the degenerate main metric in
${\cal S}^{0[\w_2]+,+}$ is: 
\be
(ds)^{(2)} =\frac {(1+\w_2\| \>\eta\| ^2)\, 
\| d\>\eta\| ^2 -\w_2
\langle \>\eta | d\>\eta\rangle^2}
{(1+\w_2\| \>\eta\| ^2)^2} 
\ee
\be
\begin{array}{l}
\| \>\eta\| ^2 =(\eta^1)^2+(\eta^2)^2+(\eta^3)^2 \qquad
\| d\>\eta\| ^2 =(d\eta^1)^2+(d\eta^2)^2+(d\eta^3)^2 \cr
\langle \>\eta | d\>\eta\rangle^2 =\eta^1\,d\eta^1+\eta^2\,d\eta^2+\eta^3 
\,d\eta^3  
\end{array}
\ee
which corresponds to the fact that in these cases three-velocity space is a
rank-one space of constant curvature $\w_2$. The hyperbolic nature of the
velocity space in relativity  has been known since a long time. For the
Galilean case 
${\cal S}^{0[0]+,+}$ the metric in phase space degenerates into a metric on
three-velocity space, which is a flat Euclidean space, with a simpler
expression:
\be
(ds)^{(2)}=(d\eta^1)^2+(d\eta^2)^2+(d\eta^3)^2 .
\ee
Notice that in both cases the interpretation is the same: distance in phase
space comes exclusively from the distance in the base  `velocity space',
where it reduces to the relative speed between two free movements. In these
three
$\w_1=0$ cases there is an invariant foliation, their leaves being identified
with the Euclidean 3d position space; the subsidiary metric 
$g_2^{(2)}$ defined in each leaf can be obtained from (\ref{zh}).

However, when $\w_1 \neq 0$ (either in NH or in DeSitter spacetimes), the
invariant foliation of the phase space disappears, and the possibility of
using a `reduced' three dimensional velocity space does no longer exists:
phase space should be approached as the six-dimensional space it is. 

The value of the second constant $\w_2$ determines another aspect of the
geometrical nature of the phase space: its curvature. The three
non-relativistic phase spaces (where $\w_2=0$), i.e., oscillating NH  
${\cal S}^{+[0]+,\dots,+}$, Galilei ${\cal S}^{0[0]+,\dots,+}$ and expanding
NH  ${\cal S}^{-[0]+,\dots,+}$ have {\it zero} curvature (compare with their
spacetime version). This means that these phase spaces are six-dimensional
{\it affine} spaces. However, the three relativistic phase spaces
(anti-DeSitter ${\cal S}^{-[-1/c^2]+,\dots,+}$, Minkowskian
${\cal S}^{0[-1/c^2]+,\dots,+}$ and DeSitter ${\cal S}^{+[-1/c^2]+,\dots,+}$
have negative curvature; only the Minkowskian phase space has an invariant
foliation with a three-dimensional base space of negative curvature. At this
point we sum up and display the results in   Table II,  which is laid
similarly to Table I to make the comparation easy. 

\medskip\medskip

{\footnotesize

 \noindent
{{\bf Table II.} The $2(N-1)$-dimensional rank-two spaces ${\cal
S}^{\w_1[\w_2]+,\dots, + }$.}
\smallskip

\noindent 
\begin{tabular}{lll}
\hline
\hline
Elliptic line-space &Euclidean line-space &Hyperbolic line-space \\
${\cal S}^{+[+]+,\dots, +}$ &
${\cal S}^{0[+]+,\dots, +}$ &
${\cal S}^{-[+]+,\dots, +}$ \\
$SO(N\!+\!1)/ SO(2) \!\otimes\!   SO(N\!-\!1)  \back $&
$ISO(N)/ \R \!\otimes\!    SO(N\!-\!1)  \back $ &
$SO(N,1)/ SO(1,1) \!\otimes\!    SO(N\!-\!1)  \back $ \\ 
Positive Curvature & 
Positive  Curvature & 
Positive  Curvature \\ 
No Invariant Foliation &
Invariant Foliation &
No Invariant Foliation \\
\quad &
Base ${\cal S}^{[+]+,+}$, Fiber ${\cal S}^{[0]+,+}$ & 
\quad \\
\hline
Oscillating NH Phase Space&Galilean Phase Space&Expanding NH Phase Space\\
${\cal S}^{+[0]+,\dots, +} $ &
${\cal S}^{0[0]+,\dots, +} $ &
${\cal S}^{-[0]+,\dots, +} $ \\
$ONH/ SO(2)  \!\otimes\!   SO(N\!-\!1)  \back $&
$IISO(N\!-\!1)/ \R \!\otimes\!  SO(N\!-\!1)  \back $&
$ENH/ SO(1,1)  \!\otimes\!    SO(N\!-\!1)  \back $ \\
Zero  Curvature & 
Zero Curvature & 
Zero  Curvature \\ 
No Invariant Foliation &
Invariant Foliation &
No Invariant Foliation \\
\quad &
Base ${\cal S}^{[0]+,+}$, Fiber ${\cal S}^{[0]+,+}$ & 
\quad \\
\hline
Anti-DeSitter Phase Space&Minkowskian Phase Space&DeSitter Phase Space\\
${\cal S}^{+[-]+,\dots, +}  $&
${\cal S}^{0[-]+,\dots, +}   $&
${\cal S}^{-[-]+,\dots, +}   $\\
$SO(N\!-\!1,2)/ SO(2) \!\otimes\!   SO(N\!-\!1)  \back $&
$ISO(N\!-\!1,1)/ \R  \!\otimes\!  SO(N\!-\!1)   \back $&
$SO(N,1)/ SO(1,1) \!\otimes\!   SO(N\!-\!1)  \back $ \\ 
Negative Curvature & 
Negative Curvature & 
Negative Curvature \\ 
No Invariant Foliation &
Invariant Foliation &
No Invariant Foliation \\
\quad &
Base ${\cal S}^{[-]+,+}$, Fiber ${\cal S}^{[0]+,+}$ & 
\quad \\
\hline
\hline
\end{tabular}}

\medskip
\medskip

\newpage


\bigskip\bigskip

\noindent
{\Large{{\bf Acknowledgements}}}

\bigskip

F.J.H. wishes to thank the organizers of this edition of the Workshop for
their kind invitation. This work has been partially supported by DGICYT
project PB94--1115 from the Ministerio de Educaci\'on y Ciencia de Espa\~na.


\end{document}